\documentstyle[11pt,paspconf,epsf]{article}

\begin{document}

\title{Quasar-Marked Protoclusters and Biased Galaxy Formation}
\author{S.G. Djorgovski, S.C. Odewahn, R.R. Gal, R. Brunner}
\affil{Palomar Observatory, Caltech, Pasadena, CA 91125, USA}
\author{R.R. de Carvalho}
\affil{Observatorio Nacional, CNPq, Rio de Janeiro, Brasil}

\begin{abstract}
We report on the current status of our search for protoclusters around
quasars at $z > 4$.
While the search is still very incomplete, clustered companion galaxies
are found in virtually every case examined so far.
The implied comoving number densities of protogalaxies are two to four orders
of magnitude higher than expected for the general field, but are comparable
to the number densities in rich cluster cores.  The comoving densities of
star formation in these regions are also enhanced by a comparable factor.
We interpret these results as an evidence for biased galaxy formation in
the highest peaks of the primordial density field.
\end{abstract}


\section{Introduction}
Our goal is to discover and study protogalaxies (PGs) clustered with known
quasars at $z > 4$, perhaps in the cores of future rich clusters.  This work
complements the efforts by other groups to study galaxy formation in the
general field, at the epochs when the universe was only a few percent of its
present age.  It also represents a powerful test of biased galaxy formation
models. 

PGs have been found using a variety of techniques: 
narrow-band Ly$\alpha$ imaging (Cowie \& Hu 1998),
Lyman-break method (Steidel {\it et al.} 1996),
as DLA absorbers (Djorgovski {\it et al.} 1996),
quasar companions (Hu {\it et al.} 1996, Djorgovski 1998),
gravitationally lensed objects (Franx {\it et al.} 1997),
serendipitously (Dey {\it et al.} 1998),
etc.
However, any single method has its own biases, and formative
histories of galaxies in different environments may vary substantially.  For
example, galaxies in rich clusters are likely to start forming earlier than in
the general field, and studies of galaxy formation in the field may have missed
possible rare active spots associated with rich protoclusters.

We are conducting a systematic search for clustered PGs, by using quasars at $z > 4$
as markers of the early galaxy formation sites (ostensibly protocluster cores).
The quasars themselves, selected from the DPOSS survey (Djorgovski {\it et al.} 1999; cf. also Kennefick {\it et al.} 1995), are purely incidental to this
search: they are simply used as beacons, pointing towards the possible sites 
of early, massive galaxy formation.
Our preliminary results are promising.

\section{Quasars as Markers of Early Galaxy Formation Sites}

Statistical studies by Madau {\it et al.} (1996) and others have outlined a global
history of galaxy formation in the general field, with a broad peak at 
$z \sim 2 - 3$.
Steidel {\it et al.} (1998) have shown that the apparent decline at
larger redshifts may not be real, and that considerable star formation goes
on even at $z > 4$.  Such PGs
must
be still very young, at most a few
percent of the present galaxian age, or $\sim 0.5 - 1$ Gyr, for any reasonable
cosmology. 
The current theoretical belief is that some subgalactic structures 
($M \sim 10^6 - 10^8 M_\odot$) may start forming at $z \sim 6 - 10$,
(e.g., Miralda-Escude \& Rees 1997), 
and more massive ($M \sim 10^{11} - 10^{12} M_\odot$)
PGs are expected to be very rare at $z > 4$, but there is still very little
known empirically about galaxy formation at such redshifts. 

Some of the first massive PGs may be the hosts of quasars at $z > 4$.
The comoving density of quasars tracks well the history of star formation in
galaxies, and both follow the merger rate evolution predicted by hierarchical
structure formation scenarios. 
The same kind of processes, dissipative merging and infall, may trigger both
star formation and the AGN activity. 
Most or all ellipticals and massive bulges at $z \sim 0$ seem to contain
central massive dark objects suggestive of an earlier quasar phase (Kormendy \&
Richstone 1995, and references therein), whose masses correlate with the
luminous old stellar masses of the host galaxies, suggesting that they formed
coevally. 
The activity may mask the undergoing star formation, but AGNs may still provide
useful pointers to the sites of early galaxy formation. 
High metallicities (up to $10 \times ~Z_\odot$!) observed in $z > 4$ quasars
(Hamman \& Ferland 1993) are indicative of a considerable chemical evolution
involving several generations of massive stars in a system massive enough to
retain and recycle their metals, comparable to the cores of giant ellipticals. 
Abundance patterns in the intracluster x-ray gas at lower redshifts are
suggestive of an early, rapid star formation phase at high redshifts
(Loewenstein \& Mushotzky 1996). 
Quasars at $z > 4$ may thus be situated in the progenitors of future giant
ellipticals and rich clusters. 

In general, the most massive density peaks in the early universe are likely to
be strongly clustered (Kaiser 1984; Efstathiou \& Rees 1988) and thus the first
galaxies may be forming in the cores of future rich clusters: the early
formation of galaxies should be closely related to the primordial large-scale
structure.  This is a generic expectation in most models of galaxy formation. 
Such bias is already seen at $z \sim 3 - 3.5$ (Steidel {\it et al.} 1998), and
it should be even stronger at $z > 4$.  It then makes sense to look for other
galaxies, with or without AGN, forming in the immediate vicinity of known $z >
4$ quasars.  We note that the quasars themselves are purely incidental to this
project: we simply want to use them as markers of the likely galaxy formation
hotspots in the early universe.

\section{Some Preliminary Results}

This approach has been proven to work.  A Ly$\alpha$ galaxy and a dusty
companion of BR 1202--0725 at $z = 4.695$ have been discovered by several
groups. 
(Djorgovski 1995, Hu {\it et al.} 1996, Petitjean
{\it et al.} 1996), and a dusty companion object has been found in the same
field (Omont {\it et al.} 1995, Ohta {\it et al.} 1995).  
Hu \& McMahon (1996)
also found two companion galaxies in the field of BR 2237--0607 at $z = 4.55$.

We have searches to various degrees of completeness in about twenty QSO
fields so far
(Djorgovski 1998; Djorgovski {\it et al.}, in prep.).  Companion galaxies
have been found in virtually every case, despite very incomplete coverage.
They are typically located
anywhere between a few arcsec to tens of arcsec from the quasars.

Some are close projected companions near the lines of
sight to quasars, selected by simple deep $R$-band imaging and confirmed
using long-slit spectroscopy. 
This method probes $\sim 100+$ comoving kpc 
scale environments of quasar hosts, which are likely still forming
through dissipative merging of protogalactic clumps. 
 
We also select candidate PGs associated and/or clustered with quasars 
by using deep $BRI$ imaging over a field of view of several arcmin,
probing $\sim 10$ comoving Mpc ($\sim$ cluster size)
projected scales.  These imaging data are obtained at the Palomar 200-inch
and the Keck-II 10-m telescope.
This is a straightforward extension of the method employed so succesfully to
find the quasars themselves at $z > 4$ (at these redshifts, the continuum drop
is dominated by the Ly$\alpha$ forest, rather than the Lyman break, which is
used to select galaxies at $z \sim 2 - 3.5$).  
The candidates would then be followed by multislit spectroscopy at the Keck.

\begin{figure}[tbh]
\plotfiddle{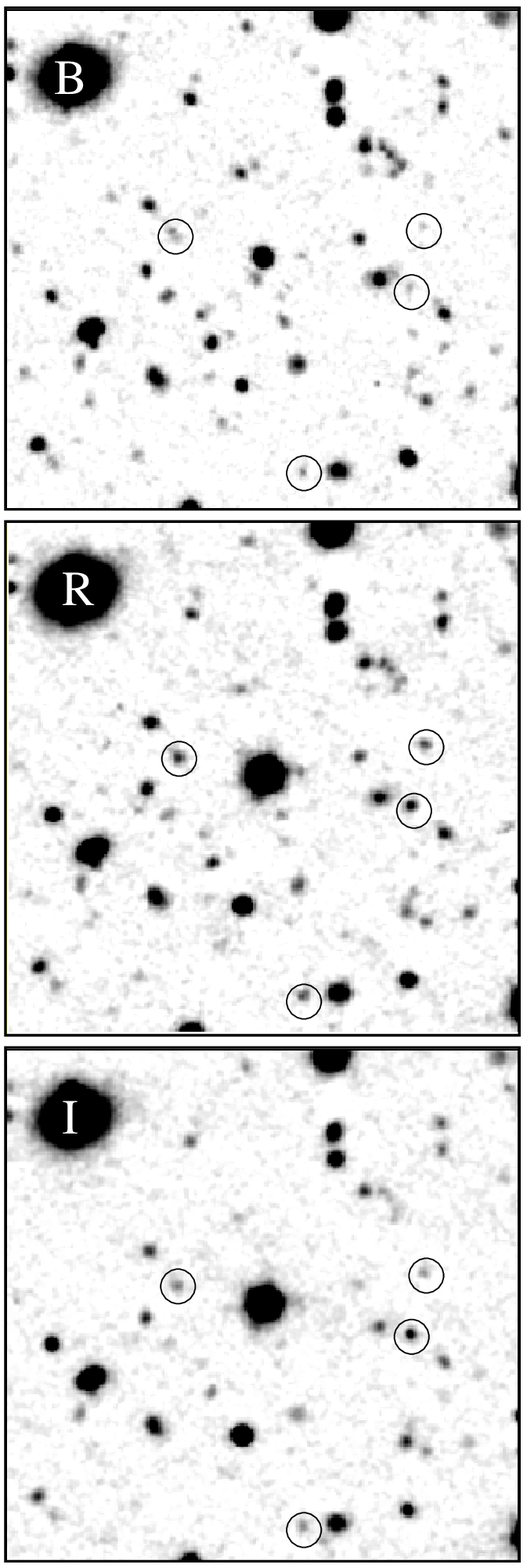}{6.0in}{0}{80}{80}{-108}{0}
\caption{
Keck images of a field centered on the quasar PSS 2155+1358, at $z =
4.26$, in $BRI$ (top to bottom).  The field shown is approximately 54
arcsec square, and is centered on the quasar.  Several color-selected
objects are circled; they are likely companion PGs clustered with the
quasar.
\label{}}
\end{figure}

As of the mid-1999, about two dozen companion galaxies have been confirmed spectroscopically, with a comparable number still in progress (the usual
reason: more
integration needed).
Their typical magnitudes are $R \sim 25^m$, implying continuum luminosities $L
\leq L_*$.  The Ly$\alpha$ line emission is relatively weak, with typical
restframe equivalent widths $\sim 20 - 30$ \AA, an order of magnitude lower
than what is seen in quasars and powerful radio galaxies, but perfectly
reasonable for the objects powered by star formation.  There are no
high-ionization lines in their spectra,
and no signs of AGN.  The SFR inferred both from the Ly$\alpha$ line,
and the UV continuum flux is typically $\sim 5 - 10 ~M_\odot$/yr, not corrected
for the extinction, and thus it could easily be a factor of 5 to 10 times
higher. 

Overall, the intrinsic properties of these quasar companion galaxies are very
similar to those of the Lyman-break selected population at $z \sim 3 - 4$,
except of course for their special environments and somewhat higher 
look-back times. 

There is a hint of a trend that the objects closer to the quasars have stronger
Ly$\alpha$ line emission, as it may be expected due to the QSO ionization
field.  
In addition to these galaxies where we actually detect (presumably
starlight) continuum, pure Ly$\alpha$ emission line nebulae are found within
$\sim 2 - 3$ arcsec for several of the quasars, with no detectable continuum at
all.  The Ly$\alpha$ fluxes are exactly what may be expected from
photoionization by the QSO, with typical $L_{Ly \alpha} \sim$ a few $\times
10^{43}$ erg/s.  They may represent ionized parts of still gaseous protogalaxy
hosts of the quasars.  We can thus see and $distinguish$ both the objects
powered by the neighboring QSO, and ``normal'' PGs in their vicinity. 

The median projected separations of these objects from the quasars are $\sim
100 h^{-1}$ comoving kpc, an order of magnitude less than the comoving r.m.s.
separation of $L_*$ galaxies today, but comparable to that in the rich cluster
cores.  The frequency of QSO companion galaxies at $z > 4$ also appears to be
an order of magnitude higher than in the comparable QSO samples at $z \sim 2
- 3$, the peak of the QSO era and the ostensible peak merging epoch.
However, interaction and merging rates are likely to be high in the densest
regions at high redshifts, which would naturally account for the propensity of
some of these early PGs to undergo a quasar phase, and to have close companions.

The implied average star formation density rate in these regions is some 2 or 3
orders of magnitude higher than expected from the limits estimated for these
redshifts by Madau {\it et al.} (1996) for $field$ galaxies, and 1 or 2 orders
of magnitude higher than the measurements by Steidel {\it et al.} at $z \sim 4$, even
if we ignore any SFR associated with the QSO hosts (which we cannot measure,
but is surely there).  These really are regions of an enhanced galaxy formation
in the early universe.

\section{Concluding Comments}

Our survey is still very incomplete, but there is already a clear indication of
an ``excess'' of PGs in these fields.  This may be an observable manifestation
of biasing, i.e., the expected clustering of the highest density peaks. 
The same explanation applies to the Steidel {\it et al.} redshift
``spikes'' at $z \sim 3 - 3.5$; what we are looking for are even denser, and
thus much rarer peaks at $z > 4$.  Because they are rare, we use quasars as
markers of sites where some structure is already forming, in order to increase
our chances.  A ``pure deep field'' approach at these redshifts is 
much harder and would likely yield fewer objects than at $z \sim 3$. 
These quasar fields are obviously very special spots in the early universe, and
they present a great opportunity to study galaxy and cluster formation, in an
enviroment deliberately different from the general field. 

It is also worth noting that (perhaps coincidentally) the observed comoving number
density of quasars
at $z > 4$ is roughly comparable to the comoving density of very 
rich clusters of
galaxies today.  Of course, there must be some protoclusters without
observable quasars in them, and some where more than one AGN is present.

Finally, there is some evidence for the clustering of quasars themselves, on
scales $\sim 100 ~h^{-1}$ comoving Mpc (cf. Djorgovski 1998).  This is an
interesting scale, corresponding to the excess power seen in some of the
redshift surveys (e.g., Broadhurst {\it et al.} 1990; Landy {\it et al.} 1996)
and also close to that of the first Doppler peak in the CMBR.
This effect may be
an artifact of a variable depth of the survey (something that we can and 
will test in the future).  It may also be due, e.g., to a patchy gravitational
lensing magnification by the foreground large-scale structure; again, we can 
test this hypothesis using the DPOSS galaxy counts.  Finally, the effect may
be due to a genuine (super)clustering of the highest peaks of the primordial
density field, some of which happen to be decorated with luminous quasars.
This would be a natural, albeit remarkable, extension of the basic concept
of biased structure formation in the early universe.

\acknowledgments

This work was supported in part by the Norris Foundation and by the Bressler
Foundation.  We thank the staff of Palomar and Keck observatories for their
expert assistance during our observing runs.

\end{document}